# BINDING ENERGY OF IMPURITY IN A SIZE QUANTIZED COATED SEMICONDUCTOR WIRE: ROLE OF THE DIELECTRIC-CONSTANT MISMATCH

Mher M. Aghasyan[*] and Albert A. Kirakosyan

Department of Solid State Physics, Yerevan State University, Yerevan 375049, Armenia

Al. Manookian 1, Yerevan, Republic of Armenia

**Abstract**

Within the framework of staircase infinitely deep (SIW) potential well model the effect of dielectric constant mismatch between the size-quantized semiconducting wire, coating and surrounding environment on impurity binding energy is considered. Calculations are done in both the absence and presence of magnetic field applied along the wire axis. By the variation method the dependences of binding energy of hydrogen-like impurity located on the wire axis on the alloy concentration, effective mass ratio, dielectric constant mismatch and magnetic field are found for the $GaAs$-$Ga_{1-x}Al_xAs$ system.

PACS Classification: 72.15.Nj; 72.90+y; 73.40.Gk

*Keywords*: Quantum well wire, nanostructure, binding energy.

[*]Corresponding author. Tel.: 3742-462-018; e-mail: aghasyan@www.physdep.r.am

## 1. Introduction

Many researchers are interested in quasi-one-dimensional (Q1D) and quasi-zero-dimensional (Q0D) nanoheterostructures because of the scientific aspects of the phenomena and the extraordinary possibilities of numerous applications [1-3]. The properties of these kinds of structures are due to their geometrical sizes and forms, and their component characteristics. As a consequence the electron gas topology for nanoheterostructures becomes a new degree of freedom [4-7].

In designing the semiconductor heterostructures it is important to take into account the difference of its dielectric constants (DC).

As it is known in most semiconductors the DC $c \geq 10$, therefore the Coulomb interaction of charge carriers with the impurity center is reduced, and the energy of binding states has characteristic values of about a few meV.

In low-dimensional systems the interaction between charged particles increases with the decreasing of characteristic dimensions of the system, since the field of charges in the surrounding medium begins to play a marked role.

If the DC of the environment is less than the DC of the system, the interaction becomes stronger than in the uniform system. In the quasi-two-dimensional system size quantized semiconductor layer, the mismatch of the DC-s is taken into account in the works [8-11].

Many papers are devoted to the study of impurity states in Q1D nanoheterosystems, but in the most of them the DC mismatch inside and outside the quantum well wires is neglected [4,5,7,12]. The effect of DC mismatch *GaAs* rectangular quantum wire surrounded by $Ga_{1-x}Al_xAs$ on electron and shallow donor impurity states in the case of both finite and infinite potential barier is studied in [13].

In this paper we report the calculation of the interaction of two charged particles in the circular semiconductor wire with the coating in the environment surrounding the system. The calculations are done with regard to DC mismatch of the wire, coating and surrounding medium and its effect on binding energy of the impurity center located on the wire axis is considered. The effect of the magnetic field applied in direction of the wire axis on binding energy with presence of DC mismatch is studied as well.

## 2. Impurity center potential

Consider the system consisting of the semiconducting wire of radius $R_1$, with the DC $c_1$, having the coating of radius $R_2$ and the DC $c_2$, immersed in the infinite environment with the DC $c_3$ (Fig. 1, a).

Solving Poisson's equation for the above mentioned nonhomogeneous system taking into account standard continuity conditions on the interfaces of "wire-coating" ($r = R_1$) and "coating-environment" ($r = R_2$) we get the expression for the potential of impurity center as:

$$j(r,z) = \frac{2e}{pc_1} \int_0^\infty dt \cos(tz) \begin{cases} K_0(tr) + N_1 I_0(tr); & r < R_1, \\ N_2[K_0(tr) + N_3 I_0(tr)]; & R_1 \leq r \leq R_2, \\ N_4 K_0(tr); & r > R_2, \end{cases} \quad (1)$$

where $I_m$ and $K_m$ are the modified Bessel functions of the second and third kinds of $m$-th order, respectively.

$$N_1 = \frac{g_1 K_1(tR_1)A_3 + K_0(tR_1)A_2}{g_1 I_1(tR_1)A_3 - I_0(tR_1)A_2}, \quad N_2 = \frac{g_1 A_1}{tR_1[g_1 I_1(tR_1)A_3 - I_0(tR_1)A_2]}, \quad (2)$$

$$N_3 = \frac{1}{A_1}(g_2 - 1)K_1(tR_2)K_0(tR_2), \quad N_4 = \frac{g_1 g_2}{t^2 R_1 R_2[g_1 I_1(tR_1)A_3 - I_0(tR_1)A_2]}, \quad (3)$$

$$A_1 = g_2 I_1(tR_2)K_0(tR_2) + I_0(tR_2)K_1(tR_2), \tag{4}$$

$$A_2 = (g_2 - 1)K_1(tR_2)K_0(tR_2)I_1(tR_1) - K_1(tR_1)A_1, \tag{5}$$

$$A_3 = (g_2 - 1)K_1(tR_2)K_0(tR_2)I_0(tR_1) + K_0(tR_1)A_1, \tag{6}$$

$$g_1 = \frac{c_1}{c_2}, \quad g_2 = \frac{c_2}{c_3}. \tag{7}$$

In the case of a homogeneous system ($c_1 = c_2 = c_3$), using Eq. (2)-(7) and the equation $K_0(x)I_1(x) + K_1(x)I_0(x) = 1/x$, it can be found, that $N_1 = N_3 = 0$ and $N_2 = N_4 = 1$. From Eq.(1) follows the well known expression for the potential in the environment for each value of $r$:

$$j(r,z) = \frac{2e}{pc_1} \int_0^\infty dt \cos(tz) K_0(tr) = \frac{e}{c_1 \sqrt{r^2 + z^2}}. \tag{8}$$

Note, that to the three different cases: $c_1 = c_2 \neq c_3$, $c_1 \neq c_2 = c_3$, $c_1 \neq c_2 \neq c_3$ at $R_2 \to \infty$, corresponds the same physical situation: wire in the infinite environment with a different DC. In the above considered cases, from Eq. (1)-(7) follow the known expressions for the potential of point charge in the wire ($r \leq R$) and surrounding environment ($R \leq r < \infty$) (see, e.g. [14]).

3. **Binding energy calculation**

In the considered system, within the framework of the SIW model the potential energy of an electron is of the form (Fig. 1, b):

$$V(\mathbf{r}) = \begin{cases} 0, & r < R_1, \\ V_0, & R_1 \leq r \leq R_2, \\ \infty, & r > R_2, \end{cases} \tag{9}$$

where $V_0$ is the value of the potential energy jump at the boundary of the wire and the coating layer. Within the framework of effective mass approximation the electron states in SIW are studied in [15].

To calculate the impurity binding energy following [7, 15], we take the ground state trial wave function in the form:

$$\Psi_0 = N \exp\left[-l\sqrt{y^2 + (z/a_B)^2}\right] \begin{cases} J_0(\boldsymbol{a}\,y), & 0 \le y < y_1, \\ C_2 I_0(\boldsymbol{b}\,y) + C_3 K_0(\boldsymbol{b}\,y), & y_1 \le y \le y_2, \\ 0, & y > y_2, \end{cases} \quad (10)$$

were $J_m$ is the Bessel function of the first kind, m-th order, $l$ is the variational parameter, $y = r/a_B$, $y_1 = R_1/a_B$, $y_2 = R_2/a_B$, $\boldsymbol{a} = \sqrt{e_{10}/E_R}$, $\boldsymbol{b} = \sqrt{m_2(v_0 - e_{10})/m_1}$, $e_{10}$ is the electron ground state energy in the SIW without impurity, $m_i$ is the effective mass of an electron in the wire ($i = 1$) and in the coating ($i = 2$), $v_0 = V_0/E_R$, $a_B = \hbar^2 c_1/m_1 e^2$ and $E_R = m_1 e^4/2\hbar^2 c_1^2$ are the effective Bohr radius and effective Rydberg energy in the wire respectively, $N$ is the normalization constant,

$$C_2 = \frac{J_m(\boldsymbol{a}_{nm} y_1) K_m(\boldsymbol{b}_{nm} y_2)}{I_m(\boldsymbol{b}_{nm} y_1) K_m(\boldsymbol{b}_{nm} y_2) - I_m(\boldsymbol{b}_{nm} y_2) K_m(\boldsymbol{b}_{nm} y_1)}, \quad (11)$$

$$C_3 = -\frac{J_m(\boldsymbol{a}_{nm} y_1) I_m(\boldsymbol{b}_{nm} y_2)}{I_m(\boldsymbol{b}_{nm} y_1) K_m(\boldsymbol{b}_{nm} y_2) - I_m(\boldsymbol{b}_{nm} y_2) K_m(\boldsymbol{b}_{nm} y_1)}. \quad (12)$$

The binding energy of the impurity is defined as the difference of the ground state energy of the system without impurity, i.e., $e_{10}$, and the ground state energy $e(y_1, y_2)$ with impurity: $e_b(y_1, y_2) = e_{10} - e(y_1, y_2)$.

Turning to the dimensionless parameters and using the expression (1) one can get for the binding energy:

$$\frac{e_b(y_1, y_2)}{E_R} = -I^2 + \frac{8}{p} I \frac{p+q}{f+g} - I \frac{(1-m_1/m_2)}{f+g}\left[J_0^2(y_1 a)K_0(2I y_1) - I g\right], \quad (13)$$

where:

$$f = \int_0^1 J_0^2(y_1 a t)K_1(2I y_1 t) t^2 dt, \quad (14)$$

$$g = \int_1^{y_2/y_1} [C_2 I_0(y_1 b t) + C_3 K_0(y_1 b t)]^2 K_1(2I y_1 t) t^2 dt, \quad (15)$$

$$p = \int_0^1 t^2 dt J_0^2(y_1 a t) \int_0^\infty \frac{K_1\left(y_1 t \sqrt{4I^2 + t^2}\right)}{\sqrt{4I^2 + t^2}} [K_0(t t y_1) + N_1 I_0(t t y_1)] dt, \quad (16)$$

$$q = \int_1^{y_2/y_1} t^2 dt [C_2 K_0(y_1 b t) + C_3 I_0(y_1 b t)]^2 \times$$

$$\times \int_0^\infty \frac{K_1\left(y_1 t \sqrt{4I^2 + t^2}\right)}{\sqrt{4I^2 + t^2}} N_2 [K_0(t t y_1) + N_3 I_0(t t y_1)] dt. \quad (17)$$

When $m_1 = m_2$, $c_1 = c_2 = c_3$ and $R_2 \to \infty$ from formula (13) follows the result of work [7]. However, if $m_1 \neq m_2$, then at $R_2 \to \infty$ from (13) we obtain an expression differing from the one in [7] by the first term in the square brackets which means the decrease of the binding energy in compared with the [7] result.

**4. Calculation of the binding energy in a magnetic field**

Electron states in SIW in the presence of magnetic field, applied along the wire axis, were examined [16], and for the eigenvalues of energy was found the expression:

$$e_{nl} = \hbar w_c \left(a_{n,|l|}^B + \frac{|l|+l+1}{2}\right) = V_0 + \hbar w_c \frac{m_1}{m_2}\left(b_{n,|l|}^B + \frac{|l|+l+1}{2}\right), \quad (18)$$

where $w_c = eH/m_1 c$ is the cyclotron frequency, $n$ and $l$ are the principal and orbital quantum numbers, respectively. The quantum numbers $a_{n,|l|}$ and $b_{n,|l|}$ are determined from the continuity condition of the logarithmic derivative of the wave functions at $r = R_1$. Following to [16, 17] we take the ground state trial wave function in the form:

$$\Psi_0^B = N^B \exp\left(-\frac{x}{2}\right)\exp\left(-I\sqrt{2\left(\frac{a_c}{a_B}\right)^2 x + \left(\frac{z}{a_B}\right)^2}\right) \times$$

$$\times \begin{cases} F(-\mathbf{a}^B, 1; x), & x < x_1, \\ C_2 F(-\mathbf{b}^B, 1; x) + C_3 U(-\mathbf{b}^B, 1; x), & x_1 \leq x \leq x_2, \\ 0, & x > x_2, \end{cases} \quad (19)$$

where $F(a,b;x)$ and $U(a,b;x)$ are the confluent hypergeometric functions [18], $N^B$ is the normalization constant, $\mathbf{a}^B \equiv \mathbf{a}_{10}^B$, $\mathbf{b}^B \equiv \mathbf{b}_{10}^B$ ([16]),

$$C_2^B = \frac{F(-\mathbf{a}_{n,|l|}, |l|+1; x_1)U(-\mathbf{b}_{n,|l|}, |l|+1; x_2)}{F(-\mathbf{b}_{n,|l|}, |l|+1; x_1)U(-\mathbf{b}_{n,|l|}, |l|+1; x_2) - F(-\mathbf{b}_{n,|l|}, |l|+1; x_2)U(-\mathbf{b}_{n,|l|}, |l|+1; x_1)}, \quad (20)$$

$$C_3^B = -C_2^B \frac{F(-\mathbf{b}_{n,|l|}, |l|+1; x_2)}{U(-\mathbf{b}_{n,|l|}, |l|+1; x_2)}, \quad (21)$$

$$x = \frac{r^2}{2a_c^2}, \quad x_1 = \frac{R_1^2}{2a_c^2}, \quad x_2 = \frac{R_2^2}{2a_c^2}, \quad (22)$$

where $a_c = (\hbar c / eH)^{1/2}$ is the magnetic length. Using expression (1), after some transformations we get for the binding energy:

$$\frac{\mathbf{e}_b}{E_R} = -I^2 + \frac{8}{\mathbf{p}} I \frac{p^B + q^B}{f^B + g^B} - I \frac{(1 - m_1/m_2)}{f^B + g^B}\left[y_1^2 e^{-x_1^2} F^2(-\mathbf{a}^B, 1; x_1)K_0(2I y_1) - I g^B\right]. \quad (23)$$

with the following notations:

$$f^B = \int_0^{y_1} e^{-x} F^2(-\mathbf{a}^B, 1; x)K_1(2Iy)y^2 dy, \quad (24)$$

$$g^B = \int_{y_1}^{y_2} \left[C_2^B F(-\mathbf{b}^B, 1; x) + C_3^B U(-\mathbf{b}^B, 1; x)\right]^2 K_1(2Iy)y^2 dy, \quad (25)$$

$$p^B = \int_0^{y_1} y^2 dy e^{-x} F^2(-\mathbf{a}^B,1;x) \int_0^\infty \frac{K_1(y\sqrt{4\mathbf{l}^2+t^2})}{\sqrt{4\mathbf{l}^2+t^2}} [K_0(ty) + N_1 I_0(ty)] dt, \qquad (26)$$

$$q^B = \int_{y_1}^{y_2} y^2 dy [C_2^B F(-\mathbf{b}^B,1;x) + C_3^B U(-\mathbf{b}^B,1;x)]^2 \times$$

$$\times \int_0^\infty \frac{K_1(y\sqrt{4\mathbf{l}^2+t^2})}{\sqrt{4\mathbf{l}^2+t^2}} N_2 [K_0(ty) + N_3 I_0(ty)] dt. \qquad (27)$$

## 5. Discussion of results

The expressions, obtained in parts 2-4 generally solve the problem of finding the effect of dielectric constants of wire, coating and surrounding environment, and alloy concentration and magnetic field on binding energy of the impurity center. In numerical calculations carried out for the $GaAs$ wire coated by $Ga_{1-x}Al_xAs$ the following values of parameters have been used ([19]): $E_R \approx 5.2$meV, $a_B = 104$Å, $m_1 = 0.067 m_0$, $m_2 = (0.067 + 0.083x)m_0$ ($m_0$ is the free electron mass) and $V_0 = 1.247 x Q_e$ eV ($Q_e = 0.6$ is the conduction-band discontinuity fraction) for the concentration $x$ within the limits $0 \le x \le 0.45$. Note that in the calculations we neglect the role of the Г-Х mixing, which in the $GaAs$-$Ga_{1-x}Al_xAs$ systems begins to play a decisive role for the values $R < 50$ Å and $x > 0.5$.

In Fig. 2 the impurity binding energy dependence on the wire radius for various values of alloy concentration $x$ and DC $c_2$ and $c_3$ is presented for $R_2 = a_B$. From the comparison of curve groups 1, 2, 5 and 3, 4, 6 it follows, that as the alloy concentration $x$ increases the maximums of the curves shift to the wire axis. The increase of binding energy is conditioned by the decrease of the electron localization region in consequence of increasing potential barrier height at the border of the wire and the coating, and the strengthening of the system inhomogeneity in consequence of coating and surrounding environment DC changes.

The curves 1 ($x = 0,1$) and 3 ($x = 0,3$) correspond to the model calculations for a fully uniform system with the DC $c = 13,18$.

From comparison of curves 2 and 4 it follows that at $c_3 = 10,06$ ($AlAs$) the change of alloy concentration from 0,1 to 0,3 (the DC of coating decreases by about 5%) increases the binding energy by 27%.

Because of small linear dimensions of the system ($R_1 \leq R_2 \sim a_B$), the impurity center field is concentrated out of the wire, essentially in the surrounding environment, so the DC changes of environment have a considerable effect on binding energy. Indeed, from the comparison of curves 1 and 2 follows, that at $x = 0,1$ $\Delta c_2 = 0,312$, $\Delta c_3 = 3,12$, and the relative change of binding energy is about 14%. At $\Delta c_2 = 0,936$, $\Delta c_3 = 3,12$ (curves 3 and 4), this change is about 16%. So, one can get that relative change of binding energy corresponding to the coating DC change $\Delta \tilde{c}_2 = 0,624$ with width $R_2 - R_{1,max} \approx 80 \text{Å}$, is about 2%. If the system is in the vacuum (curves 2 and 5), then $\Delta c_3 = 9,06$ and the relative change of binding energy at $x = 0,1$ equals 2,8, and at $x = 0,3$ (curves 4 and 6) is about 2,24. Note, that the decreasing of the relative change of binding energy is conditioned by the electron removal to the wire axis, at the same time, away from the environment border.

With the wire radius increasing the effect of the DC mismatch on binding energy decreases, and the minimums at $R_1 \approx 0,9 a_B$, caused with the effective mass mismatch ([15]), are smoothed (the more is the inhomogeneity, as smoothed the minimums).

The dependence of binding energy $e_b$ on coating radius $R_2$ is examined, too. For a fixed value of $R_1$, $e_b(R_1, R_2)$ has a maximum at $R_2 = R_1$ and then falls abruptly, tending to the value at $R_2 \to \infty$. The curves corresponding to large values of $x$, and bigger values of $c_2$ and $c_3$ (at fixed $x$) decrease relatively slowly.

In Fig. 3 the impurity center binding energy dependence on the wire radius is shown for the wire in an infinite environment ($R_2 \to \infty$ curves 1 and 2) and for SIW at $R_2 = a_B$ (curves 3 and 4) at $x = 0,3$. Curves 1 and 3 correspond to the model calculation with $c_1 = c_2 = 13,18$ and $c_1 = c_2 = c_3 = 13,18$ respectively. As one can see from fig. 3, when we take into account wire and coating DC mismatch ($c_2 < c_1$) the binding energy in SIW (curve 4) is always greater than for the wire in the infinite environment (with $c_2$, curve 2). But with the wire radius increasing the role of DC mismatch decreases faster for the wire in the infinite environment than in SIW (curves 2 and 1).

In Fig. 4 the impurity center binding energy dependence on the wire radius is presented for various values of magnetic field at $x = 0.3$, $R_2 = a_B$, $m_2 \ne m_1$. With the increase of $B$ the binding energy increases quickly within the region $R_1 \le 0.1a_B$ and $R_1 \ge 0.6a_B$ and slowly in the region $0.2a_B \le R_1 \le 0.5a_B$. Such behaviour of binding energy is the consequence of the fact that at $0.2a_B \le R_1 \le 0.5a_B$ the size-quantization prevails the magnetic one. As was noted above (fig. 2), the minimums at $R_1 \approx 0,9a_B$, caused by the effective mass mismatch in the wire and coating, are smoothed because of the systems inhomogeneity. With the increase of $B$ the depth of minimum becomes smaller and tends to zero because the strong field localizes electron in the near axis region. As one can see, this minimums completely vanish for the heterostructure $GaAs\text{-}Ga_{0,7}Al_{0,3}As$-vacuum at $B = 25\,T$, and for $GaAs\text{-}$

$Ga_{0.7}Al_{0.3}As$ - $AlAs$ at $B > 50$ T, since the impurity field in the main is in the surrounding medium because of the small sizes of the system.

For the fixed value of $R_1$ $e_b$ has a maximum at $R_2 = R_1$ and then falls abruptly, tending to the limit as $R_2 \to \infty$. The curves corresponding to the large values of $B$ and $c_2$, $c_3$ (for fixed $x$) decrease relatively quickly.

In Fig. 5 the impurity center binding energy dependence on magnetic field is presented for various parameters values of the problem. From comparison of the curves 1 ($x = 0,1$) and 2 ($x = 0,3$), to which corresponds the model calculation $c_2 = c_3 = 13,18$, and for 3 and 4, 5 and 6, follows that the $e_b$ rise velocity depending on $B$, decreases with the alloy concentration rise, owing to the increase of the potential barier on the border between the wire and the coating. Although with the increase of $B$ the electron localization radius in the wire axis region decreases, at the fixed $x$, with the decreasing of $c_2$ (curves 1 and 3, 2 and 4) and $c_3$ (curves 3 and 5, 4 and 6) the rise velocity $e_b$ increases depending on $B$. This is the consequence of field concentration in the main in surrounding medium.

## 6. Conclusion

Thus, according to the obtained results, the regard of wire, coating and environment DC mismatch can appreciably effect on the binding energy of the impurity center, and this effect increases with the rise of system inhomogeneity, caused both by the increase of alloy concentration and the decrease of the DC of coating and environment. When we take into account the DC mismatch, the binding energy is higher in the SIW, than in the infinite environment surrounding the wire, however the last system is more sensible to inhomogeneity.

The presence of a magnetic field leads to a rise of the binding energy, at that its velocity rise depending on magnetic field increases both with the decreasing alloy concentration and the decreasing DC of coating and environment.

According to the obtained results one can ascertain, that the neglecting of dielectric mismatch of the system in calculating binding energy leads to considerable errors, especially when the wire radius is decreasing and the alloy concentration is increasing.

**REFERENCES**

[1] Optical Properties of Semiconductor Quantum Dots, ed. by U. Woggon, Springer-Verlag, Heidelberg, 1997.

[2] N.V. Tkach, I.V. Pronishin, A.M. Makhanec, Fiz. Tv. Tela (in russian), 40, 557 (1998).

[3] R.R.L. De Carvalho, J.R. Filho, G.A. Farias, and V.N. Fieire, Superlattices and Microstructures, 25, 221, (1999).

[4] C.L. Foden, M.L. Leadbeater, J.H. Burroughes, and M. Pepper. J. Phys. Cond. Mat., 6, L127 (1994).

[5] C.L. Foden, M.L. Leadbeater, and M. Pepper. Phys. Rev. B, 52, R8646 (1995).

[6] Jeongnim Kim, Lin-Wang Wang, and A. Zunger. Phys. Rev. B, 56, R15541 (1997).

[7] J.W. Brown and H.N. Spector. J. Appl. Phys., 59, 1179 (1986).

[8] N.S. Rytova, Vestnik MGU, 3, 30 (1967).

[9] L.V. Keldysh, Pisma Zh. Eksp. teor. fiz. (JETP Lett.) 29, 716, (1979).

[10] S. Fraizzoli, F. Bassani and R. Buczko. Phys. Rev. B, 41, 5096, (1990).

[11] J. Cen, K.K. Bajaj, Phys. Rev. B, 48, 8061 (1993).

[12] F.A.P. Osorio, M.H. Degani, and O. Hipolito. Phys. Rev. B, 37, 1402 (1988).

[13] Zhen-Yan Deng, and Shi-Wei Gu, Phys. Rev. B, 48, 8083, (1993).

[14] D.D. Ivanenko, A.A. Sokolov, Classical Field Theory, Moscow&Leningrad, 1951.

[15] M.M. Aghasyan and A.A. Kirakosyan, J. Contemp. Phys. (Armenian Ac. Sci), vol. 34, p.17, 1999.

[16] M.M. Aghasyan and A.A. Kirakosyan, J. Contemp. Phys. (Armenian Ac. Sci),

vol. 34, p.154, 1999.

[17] S.V. Branis, Gang Li, and K.K. Bajaj. Phys. Rev. B, 47, 1316 (1993).

[18] H.Bateman and A. Erdelyi, Higher Transcendental Functions, (in russian) Nauka, 1974.

[19] S. Adachi. J. Appl. Phys., 58, R1 (1985).

[20] S. Pescetelli, A.Di Carlo, and P.Lugli. Phys. Rev., B, 56, R1668 (1997).

Fig. 1.

Fig. 2. The impurity center binding energy dependence on wire radius ($R_2 = a_B$, $c_1 = 13{,}18$).

Fig. 3. The impurity center binding energy dependence on the wire radius is shown for the wire in an infinite environment ($R_2 \to \infty$ curves 1 and 2) and for SIW at $R_2 = a_B$ (curves 3 and 4).

Fig. 4. The Impurity center binding energy dependence on the wire radius for various values of magnetic field at $x = 0.3$, $R_2 = a_B$.

Fig. 5. The impurity center binding energy dependence on magnetic field for various parameters of problem: $R_1 = 0{,}75 a_B$, $R_2 = 1{,}5 a_B$, $c_1 = 13{,}18$:

$x = 0{,}1$: 1. $c_2 = c_3 = 13{,}18$, 3. $c_2 = 12{,}868$, $c_3 = 10{,}06$, 5. $c_2 = 12{,}868$, $c_3 = 1$:

$x = 0{,}3$: 2. $c_2 = c_3 = 13{,}18$, 4. $c_2 = 12{,}244$, $c_3 = 10{,}06$, 6. $c_2 = 12{,}244$, $c_3 = 1$.

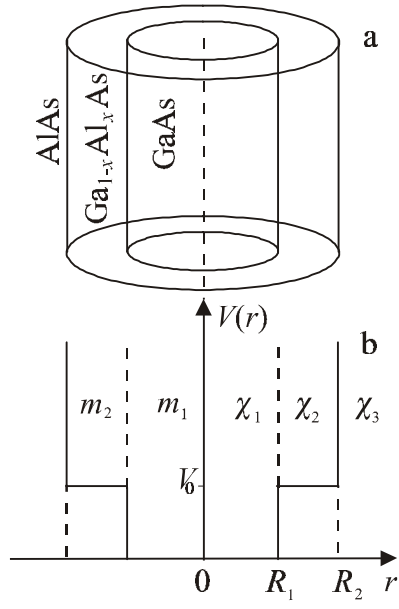

Fig. 1.

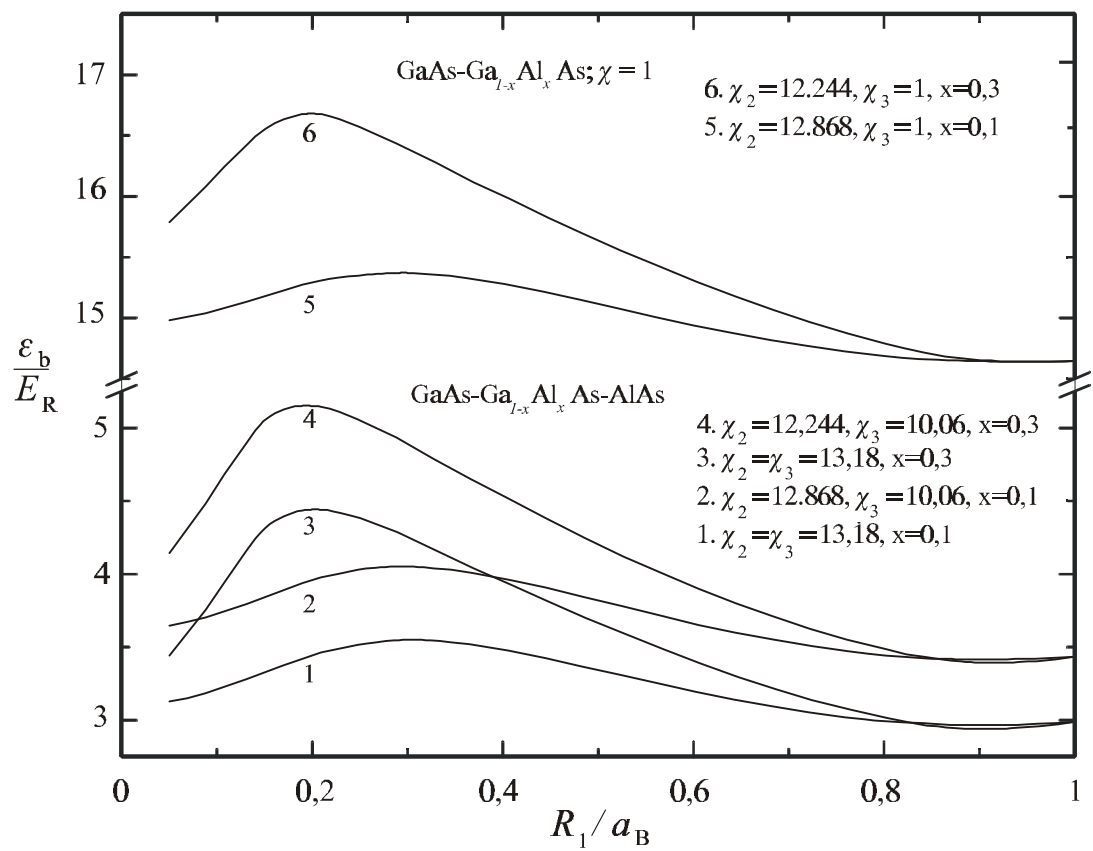

Fig. 2.

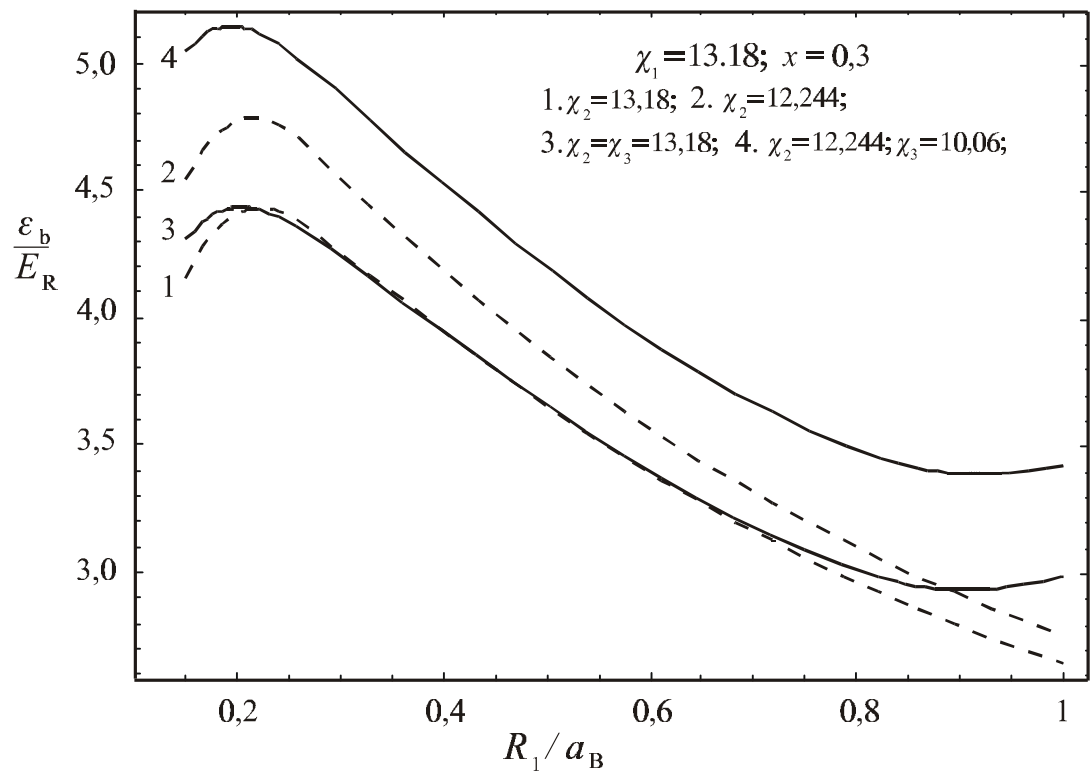

Fig. 3.

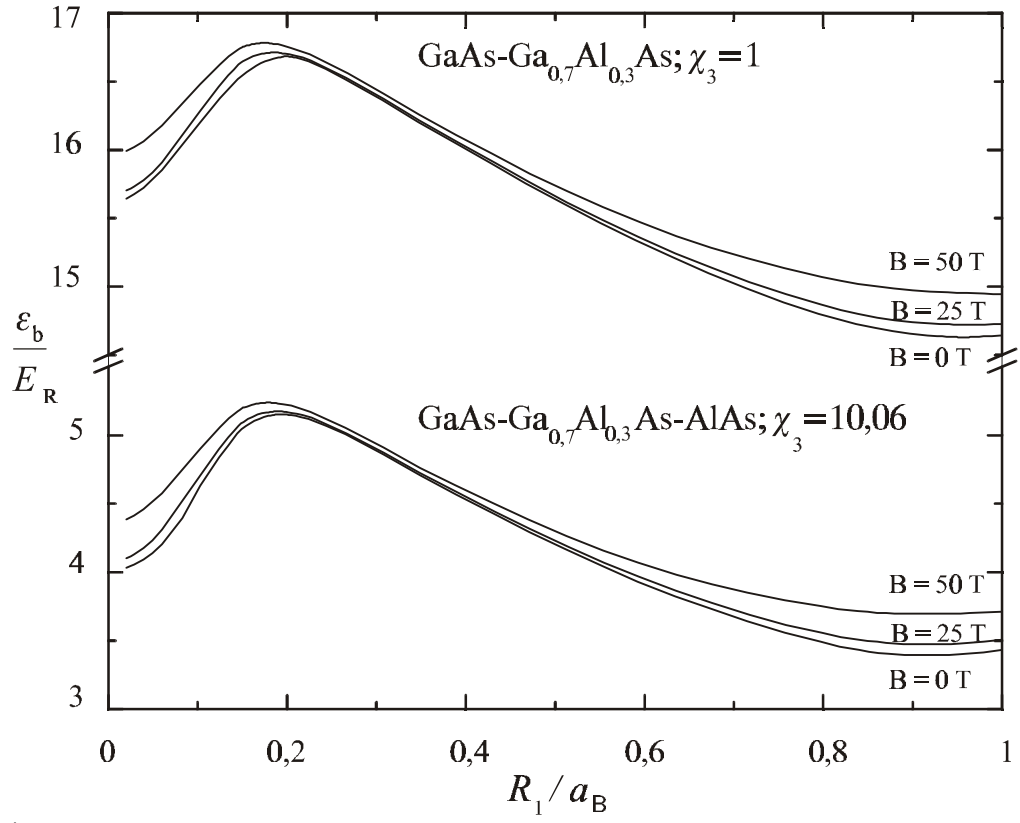

Fig. 4.

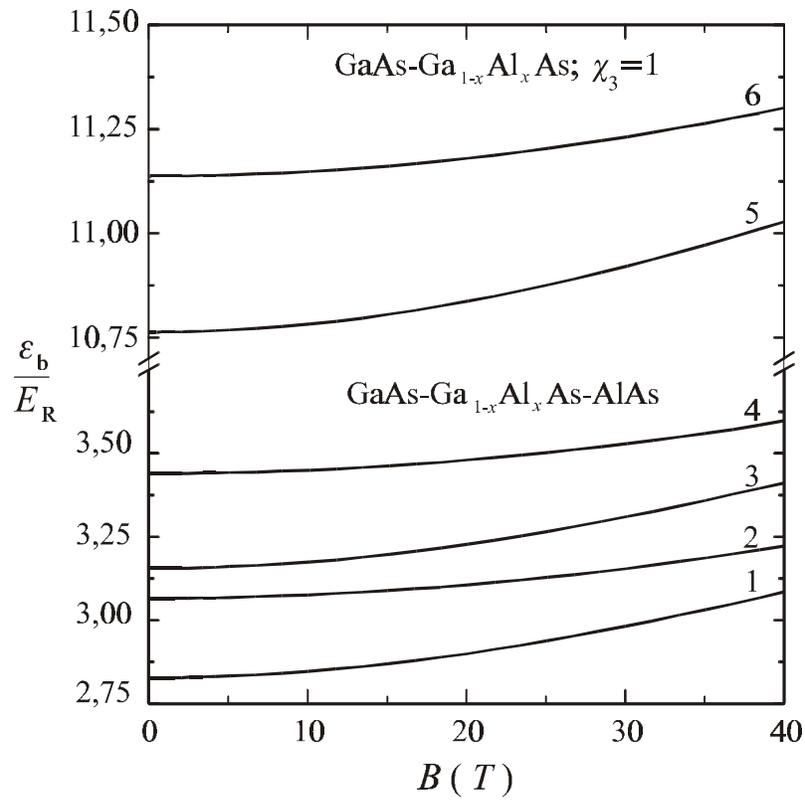

Fig. 5.